
%
%
%
  \MAINTITLE={
Carbon abundances in F and G dwarfs\FOOTNOTE {Based on observations
carried out at the Euro\-pean Sou\-thern Observatory, La Silla, Chile}
 }
  \AUTHOR={ Helena Andersson and Bengt Edvardsson }
  \INSTITUTE={ Uppsala Astronomical Observatory, Box 515, S-751 20 ~Uppsala,
Sweden }
  \ABSTRACT={
We have determined carbon abundances or upper limits from the forbidden
[C\,{\sc i}] line at 8727.13\,\AA ~for 85 F and early G type main sequence
stars, with metallicities ranging between ${\rm [Fe/H]}=-1.0$
and ${\rm [Fe/H]}=+0.25$.
The [C\,{\sc i}] line has not been previously used for a study of this size.
We find that the C/Fe abundance ratio is slowly decreasing with time and
increasing metallicity in the disk, but with considerable observational
scatter.
A discussion of results in the literature supports this result.
Our data do not reveal any gradients with galactocentric birth distance for our
stars, but more accurate observations are desirable since possible gradients
should be important to the question of the synthesis of C and Fe in the disk.
}
  \KEYWORDS={
 Galaxy: abundances
 -- Galaxy: evolution
 -- stars: abundances }
  \THESAURUS={7(10.01.1; 10.05.1; 08.01.1;)}
\OFFPRINTS={Bengt Edvardsson}
\DATE={Received 15 December 1993, accepted 17 May 1994}
\maketitle
\MAINTITLERUNNINGHEAD{Carbon abundances in F and G dwarfs }
\AUTHORRUNNINGHEAD{Andersson \& Edvardsson }
\titlea {Introduction}
Carbon is one of the most abundant elements in the Galaxy and plays an
important r\^ole for stellar interior opacities and energy generation.
Clearly it is of vital interest to have knowledge of stellar carbon abundances
since it tells us about the evolution of a key chemical element in the Galaxy.
Studies of carbon abundances
in dwarf stars have been made before, for example the study made by
Laird (1985) of 116 dwarfs, using CH lines, claims that a constant
[C/Fe] ratio is seen over the metallicity range from $-2.45$ to $+0.50$.
A study was also made by Clegg et~al. (1981) of 20 F and
G type main sequence stars with metallicities ranging from $-0.9$
to $+0.4$, where they used CH lines, high excitation C\,{\sc i} lines and,
for a few mainly metal rich stars, the [C\,{\sc i}] line at 8727\,\AA.
They concluded that the carbon abundance follows the iron abundance.
A study made by Tomkin et~al. (1986) using CH lines of stars with metallicities
ranging from $-2.50$ to $-0.70$ shows a constant [C/Fe] ratio for
metallicities down to [Fe/H]$\;=-1.8$, where an increasing [C/Fe] ratio was
found for stars of lower metallicities.
This trend is also found in a study made by Carbon et~al. (1987)
who, using CH lines, measured a practically constant [C/Fe] ratio
down to [Fe/H]$\;=-2.0$ where they found an up-turn.
A critical discussion and re-normalisation of the earlier results can
be found in the review by Wheeler et~al. (1989).

In this study we present carbon abundances and upper limits derived from
observations of the forbidden carbon line at 8727.13\,\AA ~for 85 stars
with metallicities ranging from [Fe/H]$\;=-1.0$ to
[Fe/H]$\;=+0.25$ (we adopt the square bracket notation of relative logarithmic
abundance ratios, e.g. [Fe/H]$\;= \log (N_{\rm Fe}/N_{\rm H})_{\rm star} -
\log (N_{\rm Fe}/N_{\rm H})_\odot$).
The forbidden [C\,{\sc i}] line has not previously been used to any large
extent
since it is weak and therefore hard to measure, especially in metal poor stars.
The forbidden line is an interesting abundance criterion, since it should not
be sensitive to non-LTE effects, contrary to what may be expected for other
lines.
It is also a low excitation line and therefore less sensitive to uncertainties
in the adopted model atmospheric parameters than are the frequently used high
excitation C\,{\sc i} lines and CH lines.

The observations and equivalent width measurements are described in Sect.~2.
The data is analyzed in Sect.~3 using the new line-blanketed model atmospheres
of Edvardsson et~al. (1993) and the resulting carbon abundances are used in
Sect.~4 for a statistical study of the [C/Fe] vs [Fe/H] relation.
In Sect.~5, we finally discuss our and other's results and possible
implications
for the understanding of the chemical evolution of the galactic disk.

\titlea {Observations and data reductions}
\titleb {Selection of stars, observations and data reductions}
The stars were selected from the ESO part of the large sample of F and early G
type main sequence stars studied by Edvardsson et~al. (1993, EAGLNT hereafter).
The spectral observations were originally obtained for the EAGLNT project
and the signal-to-noise ratios are not ideal for the very weak [C\,{\sc i}]
line, why the uncertainties in the individual stellar carbon abundances
presented below are significant.
The spectra of the wavelength region 8710\,\AA ~to 8780\,\AA ~were used for
measuring the 8727.13\,\AA ~[C\,{\sc i}] line.
The spectra were obtained with the ESO 1.4\,m Coud\'e Auxiliary Telescope and
the Coud\'e Echelle Spectrometer (CAT/CES), and recorded by either the Long
Camera and a Reticon detector or, for the fainter stars, the Short Camera and a
thinned and coated CCD detector (RCA SID 503).
The observations and basic data reductions are described in EAGLNT.
We estimate from the appearance of the spectra that the S/N ratios vary between
70 and 350 for the wavelength region, but that only a few spectra have
${\rm S/N} < 150$.
The CCD detector was afflicted by serious interference fringes in the
8745\,\AA ~region, which, after careful reductions, still remain with an
amplitude of up to 1\%, and a frequency of 0.7 -- 3.5\,\AA.
EAGLNT tried to estimate and reduce this effect in the data reductions, but
it may cause serious errors in our measurements of the very weak [C\,{\sc i}]
line, especially for the fainter (mainly the more metal poor) stars.

The 85 stars, for which equivalent widths or upper limits were measurable, are
given in Table~1 together with iron abundances from EAGLNT and carbon
abundances.
\begtabemptywid 19.1 cm
\tabcap{1} {
Data for the programme stars.
The equivalent widths and upper limits for the [C\,{\sc i}] line, $W_\lambda$,
are given with an error estimate when more than one spectrum could me measured.
The model parameters $T_{\rm eff}$, $\log g$, $\xi_t$, and the [Fe/H] values
are adopted from EAGLNT
}
\endtab

\titleb {Equivalent widths}
When measuring the [C\,{\sc i}] equivalent widths we found that only 37 stars
had spectra with high enough signal-to-noise ratios for the line to be measured
with some confidence.
These equivalent widths were measured with a program which adjusts a double
Gaussian to the [C\,{\sc i}] line at 8727.13\,\AA ~and a Si\,{\sc i}
line at 8728.01\,\AA.
The measurements of the equivalent widths were made with an estimated accuracy
of 25\% for the metal-rich stars and 30\% for the metal-poor stars.
The residual interference fringes in some of the spectra, discussed in
Sect.~2.1, may cause larger errors, occasionally up to a factor of 2.
Figure~1 shows the line in one metal-rich and one metal-poor star.
\begfig 8.9cm
\figure{1a-b}{
The spectra of a metal-rich star, HR\,4903, and of a metal-poor star,
HR\,8181, for which equivalent widths were measured.
Note that only the uppermost 10\% of the spectra are shown.
The wavelength scale of HR\,8181 is displaced by a radial velocity shift
}
\endfig

\begtabemptywid 21.3 cm
\tabcap{1} {(continued)}
\endtab
For the other 48 stars whose spectral S/N ratios were too poor we measured
upper limits to the equivalent widths with the help of synthetic line profiles.
Figure~2 shows two examples of spectra where only upper limits were measured.
For the star HR\,1545 the S/N ratio was so poor that no estimate of the
line strength could be made.
\begfig 9.6cm
\figure{2a-b}{
The spectra (solid line) of two stars, HR\,2530 and HR\,3578 for which
upper limits were measured, together with the synthetic line profiles (dashed
lines) used to estimate these upper limits.
The wavelength scales are displaced by radial velocity shifts
}
\endfig

\titlec {The {\rm N\,{\sc i}} 8718.83\,\AA ~absorption feature}
Measurements were also carried out on the high excitation N\,{\sc i} line at
8718.83\,\AA ~and we tried to use those to derive nitrogen abundances for our
stars.
It was found, however, that the CN line blend (cf. Lambert 1978) is totally
dominating in most of our flux spectra -- even for the Sun, whereas it gives
only a small contribution in the solar disk center spectrum used by Lambert.
We did not try to use it further.

\titlea {Determination of carbon abundances}
\titleb {Model atmospheres and fundamental parameters}
The carbon abundances have been derived using the plane-parallel,
flux-constant, LTE model atmospheres described and discussed in EAGLNT.
These models include line blanketing by
a large number of previously unconsidered weak metal lines.
For each star a specific model atmosphere was computed.
The effective temperatures, surface gravities and microturbulence
parameters were adopted from EAGLNT, while for the metallicities, we adopted
their spectroscopic values of [Fe/H].

The random errors in the model parameters were estimated by EAGLNT to be
$\pm50$\,K in $T_{\rm eff}$ and $\pm 0.07$\,dex in $\log g$.
The errors in the model metallicities and microturbulence parameters
are unimportant for the carbon and iron abundances.

The systematic errors in the effective temperatures were estimated in EAGLNT to
be in the range of $-50$\,K to $+100$\,K and in $\log g$ $\pm$0.2\,dex.

\titleb {Analysis of the {\rm [C\,{\sc i}]} line}
For each of the model atmospheres the equation of radiative transfer was solved
for a number of wavelength points in the [C\,{\sc i}] line and for the
continuum, and a theoretical equivalent width was calculated from the resulting
surface fluxes.
The carbon abundance of the model was then varied until the
calculated equivalent width agreed with the observed value.

The oscillator strength of the 1.26 eV [C\,{\sc i}] line, $\log gf = -8.21$,
was adopted from Lambert (1978) (cf. also Nussbaumer \& Rusca 1979).

The [C\,{\sc i}] line strength is, in the cool stars, affected by CO formation.
When deriving the carbon abundances we used the oxygen abundances derived
for our stars by Nissen \& Edvardsson (1992) and in EAGLNT.

According to Lambert \& Ries (1977) there is an Fe\,{\sc i} line at
8727.13\,\AA ~which will blend with the [C\,{\sc i}] line.
The atomic data for this line was taken from Kurucz \& Peytremann (1975)
except for the value of $\log gf = -3.60$ which is an upper limit taken from
Lambert \& Ries who based this value on the strengths of other
Fe\,{\sc i} lines of the same multiplet observed in the solar spectrum
(Kurucz \& Peytremann give $\log gf = -4.05$).
We have estimated upper limits of the contributions from the Fe\,{\sc i} line
to
the equivalent width for the [C\,{\sc i}] line using the upper-limit $gf$ value
of Lambert \& Ries (1977) and the iron abundances derived by EAGLNT and find
the
Fe\,{\sc i} contribution to the [C\,{\sc i}] line to be
less than 15\% for all our programme stars.

\titleb {Errors in the derived abundances}
As discussed in Sect.~2.2 the errors in the equivalent width measurements
correspond to an error of 0.11\,dex in the carbon abundance for the more
metal rich stars and an error of 0.13\,dex, or sometimes up to 0.3\,dex, for
the
more metal poor stars.
The small uncertainty in the logarithmic oscillator strength of the [C\,{\sc
i}]
line, $\pm 0.01$\,dex (Lambert 1978), gives exactly that effect on the
abundances.
The random errors in the effective temperatures and surface gravities
(see Sect.~3.1) cause very small uncertainties in [C/H] of $\pm 0.01$,
and $\pm 0.02$\,dex, respectively, and in [C/Fe] of the order of $\pm 0.02$ and
$\pm 0.02$\,dex.
With an uncertainty of $\pm 0.07$\,dex in [O/H] we get, at the most, an error
of $\pm 0.01$\,dex in the carbon abundances due to CO formation.

The systematic errors may affect the abundances in a more serious way:
Table~2 shows the changes in [C/Fe] for the three stars
HR\,784, HR\,4903 and HD\,98553 caused by changes in the model parameters
corresponding to the maximum systematic errors estimated in EAGLNT.
Obviously these errors give non-negligible uncertainties in the carbon
abundance
scale.
The forbidden, low excitation carbon line is unlikely to be affected by
non-LTE effects; we are unaware of any detailed investigations of this.

We have also searched for correlations between carbon abundance and
atmospheric parameters:
There is no correlation of [C/Fe] with either $T_{\rm eff}$ or $\log g$, while
there may possibly be a weak correlation with microturbulence parameter,
in the sense of a higher carbon abundance at lower microturbulence parameters.
We do not understand how such a correlation, if true, could occur, and we note
that the microturbulence parameters were computed in EAGLNT as a function of
$T_{\rm eff}$ and $\log g$.
\begtabfullwid
\tabcap{2} {
Changes in the derived carbon abundance scale resulting from our estimated
systematic errors in the model parameters for 3 programme stars.
($T_{\rm eff}$/$\log g$/[Fe/H]/$\xi_{\rm t}$) for the stars are, respectively:
(6287/4.37/$+0.02$/1.6), (5953/4.00/$+0.24$/1.9) and (5907/4.38/$-0.43$/1.3)
from EAGLNT.
}
\vbox {\halign {
#\hfil&\quad\hfil#&\quad\hfil#&\quad\quad\quad\hfil#&\quad\hfil#&
\quad\quad\quad\hfil#&\quad\hfil#\cr
\multispan7\hrulefill\hfil\cr
&\multispan2{\hfil ~~~HR\,784\hfil} &\multispan2{\hfil ~~~HR\,4903\hfil} &
\multispan2{\hfil ~~~HD\,98553\hfil} \cr
Uncertainty & $\Delta$[C/H] & $\Delta$[C/Fe] & $\Delta$[C/H] &
 $\Delta$[C/Fe] & $\Delta$[C/H]  & $\Delta$[C/Fe]  \cr
\multispan7\hrulefill\hfil\cr
$\Delta T_{\rm eff}=+100$\,K            & $+0.01$ & $-0.04$ &  $0.00$ &
 $-0.05$ & $+0.01$ & $-0.04$ \cr
$\Delta \log g=+0.2$                    & $+0.08$ & $+0.08$ & $+0.09$ &
 $+0.09$ & $+0.08$ & $+0.07$ \cr
$\Delta {\rm [Fe/H]}=+0.1$              & $+0.02$ & $+0.02$ & $+0.03$ &
 $+0.03$ & $+0.03$ & $+0.03$ \cr
$\Delta \xi_{\rm t}=+0.3$\,km\,s$^{-1}$ & $-0.01$ &  $0.00$ &  $0.00$ &
 $-0.01$ &  $0.00$ &  $0.00$ \cr
\multispan7\hrulefill\hfil\cr
}}
\endtab

We estimate that the total random errors in the carbon abundances are
0.10 -- 0.30\,dex, almost entirely due to equivalent width measurement errors.
The systematic effects on our abundance scale may be 0.1 -- 0.2\,dex.

\titlea {Results}
Table~1 contains the values and upper limits of [C/H] and [C/Fe] derived from
our analysis of the [C\,{\sc i}] 8727\,\AA ~line and the Fe abundances
of EAGLNT.

Our sample contains 20 stars in common with the study of Laird (1985).
The atmospheric model parameters differ rather much between the two studies,
but if Laird's results are modified to our model parameters using the
abundance differentials in Table~2 of Laird's paper we
find that the differences are, in the sense Laird's study minus ours:
$\Delta {\rm [C/Fe]} = -0.06 \pm 0.17$ (scatter standard deviation).

We also have 4 stars in common with Friel \& Boesgaard (1992).
If we similarly compare the results when their model parameters are modified to
our values we get, in the sense Friel \& Boesgaard minus ours:
$\Delta {\rm [C/Fe]} = -0.04 \pm 0.11$.

These differences are consistent with the uncertainties of the three
investigations.

\begfig  14.6cm
\figure{3a-c}{
[C/Fe] and [C/Mg] vs [Fe/H] and [C/Fe] vs logarithmic age
($\tau_9 =\;$stellar Age in Gyr) for the programme stars.
The abundance determinations are shown by filled circles and the upper
limits by open triangles
}
\endfig
Figure~3a shows [C/Fe] vs [Fe/H] for our programme stars.
The distribution of upper limits (triangles) suggests that the abundance
determinations alone (filled circles) may give an exaggerated impression of a
decrease in the C/Fe abundance ratio with metallicity.
A likely cause for this may be that the [C\,{\sc i}] line is generally
weaker and more difficult to measure in more metal poor stars (if [C/Fe] is
actually independent of metallicity).
Thus, only spectra of stars with comparatively high [C/Fe] abundances -- or in
which the noise happens to strengthen the 8727\,\AA ~feature -- yield abundance
determinations.
(Since we wish to study the carbon abundance as a function of [Fe/H] and
stellar
age, we would like to derive linear regression fits to our abundance data and
upper limits. Our attempts in this direction have failed:
Dr. E. Feigelson and colleagues (cf. La\,Valley et al. 1992, and references
therein) have kindly made available the ASURV software package, which, however,
according to the manual, is not applicable when the probability that a
measurement of an object is censored depends on the value of the censored
variable -- the carbon abundance in our case.)
It is not surprising that the scatter in the [C/Fe] abundances and upper limits
increases with decreasing metallicity; the measured equivalent width for the
8727\,\AA ~feature, on the one hand, typically changes by a factor of 2 from
the metal-poor to the metal-rich stars, while the equivalent width (assuming
conservatively [C/Fe]$\,=0$) is, in fact, expected to change by a factor of
about 5.
The problem of the CCD interference fringes in some spectra affects mainly the
metal poor stars.
Fig.~3a suggests to us that [C/Fe] may be a slightly decreasing function of
[Fe/H], ${{\Delta {\rm [C/Fe]}}/{\Delta {\rm [Fe/H]}}} \approx -0.2$\,dex.
In Fig.~3b we compare carbon and magnesium abundances.
The Mg abundances were adopted from EAGLNT who found that the abundance of Mg
and other $\alpha$ elements have decreased relative to iron
during the life of the disk.
The decrease in [C/Fe] is, however, probably not as strong at that found for
[Mg/Fe] by EAGLNT.
In Fig.~3c the stars older than 6\,Gyr show about 0.15\,dex
higher [C/Fe] ratios than the younger stars.
It is not clear from our data whether there is a sudden change at 6\,Gyr or
if there is a smooth transition.

Figure~4 shows [C/Fe] versus [Fe/H] for three different ranges in $R_{\rm m}$;
$R_{\rm m}$ being defined as $(R_{\rm p} + R_{\rm a})/2$ where $R_{\rm p}$ and
$R_{\rm a}$ are the peri- and apo-galactocentric distances for the stellar
orbits.
The values of $R_{\rm m}$ have been adopted from EAGLNT and are
estimates of the stellar birth distances from the galactic centre.
Clearly, there is no convincing difference in the [C/Fe] ratio between
the 3 different galactocentric birth-site ranges, although more accurate
abundance data would be welcome.
\begfig  13.2cm
\figure{4a-c}{
[C/Fe] vs [Fe/H] for three different $R_{\rm m}$ ranges:
$R_{\rm m}<7$\,kpc ({\bf a}), $7 \le R_{\rm m} <9$\,kpc ({\bf b})
and $R_{\rm m} \le 9$\,kpc ({\bf c}).
The Sun is now assumed to be at 8.0\,kpc and has an $R_{\rm m}$ of 8.4\,kpc
}
\endfig

\titlea {Discussion and conclusions}
\titleb {Observational results}
This work set out to investigate the history of carbon production in the
galactic disk.
We here schematically define the disk to be representatively sampled by solar
neighbourhood and open cluster dwarfs with metallicities
${\rm [Fe/H]} \ge -1.0$, and concentrate below on stars in this metallicity
range.

In Sect.~4, Fig.~3 we found that the C/Fe abundance ratio is probably slowly
decreasing with increasing [Fe/H] and with decreasing stellar age in the disk.
We could not detect any variation in [C/Fe] with galactocentric birth radius,
i.e. no radial gradient, although more accurate data are desirable.

Our result of a varying [C/Fe] ratio, based on the [C\,{\sc i}] line at
8727\,\AA, is at odds with the conclusions of Laird (1985), [C/Fe]$\;=0$ in
the disk, from an analysis of CH lines in field dwarfs.
Laird's data, however, allows a small decrease of [C/Fe] by 0.10\,dex from
[Fe/H]$\;=-1.0$ to [Fe/H]$\;=0.0$, as shown by the linear least-squares fit in
Fig.~5a.
Clegg et~al. (1981) used CH lines, high excitation C\,{\sc i} lines and, for
a few stars, the 8727\,\AA ~[C\,{\sc i}] line.
They reported a slope of [C/Fe] vs [Fe/H] of $-0.16 \pm 0.08$ and commented on
systematic differences between results from the CH and C\,{\sc i} lines.
In Fig.~5c and d we have separated Clegg's et~al. data into results from
neutral and molecular carbon lines, since their relative abundance scales may
be uncertain.
Again, a decrease in [C/Fe] may be seen for the C\,{\sc I} lines, while
for the molecular lines this is possible but very uncertain.
Slightly smaller values ([C/Fe]$\; \approx -0.2$, Fig.~5b) were derived from CH
lines in 9 disk dwarfs in a narrow range of [Fe/H] by Tomkin \& Lambert (1984).
These and other results were reviewed by Wheeler et~al. (1989), who also
discussed various systematic differences between the analyses.
Wheeler et~al. concluded that that [C/Fe]$\;=0$ in the disk and also in the
halo, at least down to [Fe/H]$\;=-2.0$, although inspection of their Fig.~1
shows [C/Fe] to be slightly decreasing with increasing [Fe/H] in the disk;
an eye estimate suggests that
${{\Delta {\rm [C/Fe]}}/{\Delta {\rm [Fe/H]}}} \approx -0.15$\,dex.
\begfig  11.2cm
\figure{5a-f}{
Previous results for disk dwarfs ([Fe/H]$\;\ge-1.0$)
Results from different carbon abundance criteria have been separated whenever
possible
}
\endfig

More recently, Friel \& Boesgaard (1992 and references therein) used Fe lines
and high excitation C\,{\sc i} lines for accurate abundance determinations in
open clusters and field stars.
Fig.~5e, which shows the results for field disk dwarfs of Friel \& Boesgaard
also suggests a slowly decreasing [C/Fe] abundance ratio with
increasing metallicity, although they caution that the high excitation
C\,{\sc i} lines in the more metal poor stars are sensitive to line blends and
to systematic errors in effective temperatures and surface gravities.

Summary: this and several previous analyses, using three different carbon
abundance criteria (with somewhat uncertain relative zero points), show that
[C/Fe] is probably slowly decreasing with time and increasing metallicity in
the galactic disk.

\titleb {Carbon in Ba dwarf stars}
Tomkin et~al. (1989) discovered a type of mildly s-element rich dwarf stars
-- Ba dwarfs.
Six of these were listed by EAGLNT and three of these are contained in our
sample: HR\,2906, HR\,5338 and HD\,6434.
Assuming that this class of stars is related to the ``classical'' Ba~II (giant)
stars and noting that Ba giants are also carbon rich (Lambert 1992), it may be
surprising that we see no sign of unusual carbon abundances:
Only HR\,2906 ($\log \tau_9=0.62$) has a carbon determination,
which is typical for other stars of similar iron abundance and age,
cf. Table~1 and Fig.~3.
HR\,5338 ($\log \tau_9=0.48$) and HD\,6434 ($\log \tau_9=1.10$) have upper
limit
carbon abundances close to the usual values for their respective metallicities
and ages.

\titleb {Carbon and iron production in the disk}
When we are confident concerning the run of [C/Fe] vs [Fe/H] in the disk,
it will have to be explained in terms of their production sites.
The answers may not be very simple: Amari et~al. (1993) found evidence from
meteoritic graphite grains that there were at least three sources of carbon,
with different ratios of $^{12}$C/$^{13}$C, contributing to the Solar nebula.
For iron it is now established that different production sites must be invoked
in the halo and in the disk to account for differences in relative abundance
ratios.

EAGLNT attributed iron production in the disk to Supernovae of Type Ia;
due to stars in the initial mass range 5 -- 7\,$M_\odot$.
Such objects are presumably due to C-deflagration explosions of C-O white
dwarfs
producing 0.5 -- 0.6\,$M_\odot$ of $^{56}$Ni which rapidly decays through
$^{56}$Co to $^{56}$Fe (Nomoto et~al. 1984).
If, in this violent process, a mass of $^{12}$C equal to about twice the mass
of
the $^{56}$Fe produced is expelled into the interstellar medium before it gets
caught by the deflagration wave, it could account for the C/Fe abundance ratio
(approximately 10 by number of atoms) observed in the Sun and in other disk
dwarfs.
This scenario would require that less $^{56}$Fe is produced than
Nomoto et~al. (1984) predict, since otherwise the Chandrasekhar mass would be
exceeded by the white dwarf.
As a balance, it should be pointed out that the white-dwarf carbon-deflagration
models studied by Nomoto et~al. produce relative numbers of $^{12}$C to
$^{56}$Fe nuclei in the range 1.5 to 0.25, ratios far too small to be important
for carbon production in the galactic disk.

Wolf-Rayet stars are initially massive stars which have lost their envelopes
through strong stellar winds and which continue to loose mass from their bare
cores at very high rates.
They may, in the WC phase, contribute a significant fraction of the
$^{12}$C observed in disk stars (Maeder \& Meynet 1993).
Their numbers are strongly increasing with increasing metallicity, and their
mass-loss rates and distribution over various sub-types change with
metallicity.
One theoretical prediction is that their carbon yields should decrease with
increasing metallicity (Maeder 1992).
Therefore we do not know whether the annual amount of $^{12}$C injected to the
ISM by WR stars is increasing or decreasing with overall metallicity.

Stars of masses in the range 1 -- 8\,$M_\odot$ also give rise red giants and
supergiants, which, via winds and planetary nebulae, add freshly synthesised
carbon to the interstellar medium.
Carbon stars are abundant in metal-poor populations like those of the
Magellanic
Clouds, and few in metal-rich populations like that of the galactic nuclear
bulge.
Their carbon dominated atmospheres are efficiently forming solid particles,
which by action of the radiation pressure give rise to strong stellar winds
with
high relative carbon abundances (cf. Olofsson et~al. 1993a,b).
They may well supply an important fraction of the carbon in the galactic disk
(Lee \& Wdowiak 1993).
If that is the case, their decreasing frequency, as a function of metallicity,
might cause a [C/Fe] ratio which decreases with time.
An interesting observation in this connection concerns
the metallicity sensitivity of the relative frequences of spectral type J and N
carbon stars, the J-types often having low $^{12}$C/$^{13}$C (down to
$\approx$3
which is yielded by equilibrium shell CNO burning) and the N-types showing
$^{12}$C/$^{13}$C ratios between 30 and 100 (cf. Gustafsson 1989).
Theoretically, Renzini \& Voli (1981) predicted larger fractions of
$^{13}$C-rich carbon stars in metal-poor environments,
as well as larger fractions of carbon stars.
On the contrary, it observationally appears as if the J-type frequency is much
less metallicity sensitive than the N-type phenomenon (Lloyd Evans 1993), so
that the mean $^{12}$C/$^{13}$C ratio in winds from carbon stars
should decrease with increasing metallicity.
The systematic investigation of $^{12}$C/$^{13}$C ratios in disk stars should
therefore give additional information concerning the production of carbon in
the
Galaxy.

Since iron and carbon in the galactic disk are supposedly formed by different
processes, having probably different metallicity sensitivities, in stars which
are unlikely to be of exactly the same mass ranges,
it is {\it not} surprising that [C/Fe] is varying with [Fe/H].
It would be important for the understanding of the nucleosynthesis
mechanisms operating in the galactic disk if one could be confident concerning
the metallicity trend in this ratio.
A good answer to this question should be able to discern a trend in
${{\Delta {\rm [C/Fe]}}/{\Delta {\rm [Fe/H]}}}$
to an accuracy of at least $\pm 0.05$.
We should also look out for trends with stellar galactocentric birth-radius,
since such trends may tell us which stellar mass-ranges are important in the
production of C and Fe.
In view of the robustness of the [C\,{\sc i}] line criterion to non-LTE effects
and effective temperatures, and the availability of high S/N, high spectral
resolution instruments, this line should be carefully observed in a good sample
of disk dwarfs in an attempt to resolve these problems.

\acknow{
We thank K.\,Eriksson and B.\,Gustafsson for essential comments and suggestions
to this work.
O.\,Morell is thanked for generous help with data reduction software and for
valuable discussions.
We have also benefited from discussions with T.\,Lloyd Evans and B.\,Westerlund
concerning the carbon star phenomenon
and with K.\,Nomoto concerning supernova nucleosynthesis.
J.\,Andersen, B.\,Gustafsson and P.\,E.\,Nissen are thanked for letting us use
their observational material to measure the weak carbon lines.
This research was supported by the Swedish Natural Sciences Research Council.
}

\begref{References}
\ref Amari S., Hoppe P., Zinner E., Lewis R.S. 1993, Nat 365, 806
\ref Carbon D.F., Barbuy B., Kraft R.P., Friel E.D., Suntzeff N.B. 1987,
PASP 99, 335
\ref Clegg R.E.S., Lambert D.L., Tomkin J. 1981, ApJ 250, 262
\ref Edvardsson B., Andersen J., Gustafsson B., Lambert D.L., Nissen P.E.,
Tomkin J. 1993, A\&A 275, 101 (EAGLNT)
\ref Friel E.D., Boesgaard A.M. 1990, ApJ 351, 480
\ref Friel E.D., Boesgaard A.M. 1992, ApJ 387, 170
\ref Gustafsson B. 1989, ARA\&A 27, 701
\ref Kurucz R.L., Peytremann E. 1975, A table of semiempirical gf values,
Smithsonian Astrophys. Obs. Special Report No 362
\ref Laird J.B. 1985, ApJ 289, 556
\ref Lee W., Wdowiak T.J. 1993, ApJ 417, L49
\ref Lambert D.L. 1978, MNRAS 182, 249
\ref Lambert D.L. 1992, in: Elements and the Cosmos, Edmunds M.G.,
Terlevich R.J. (eds.), Cambridge University Press, 92
\ref Lambert D.L., Ries L.M. 1977, ApJ 217, 508
\ref La Valley M., Isobe T., Feigelson E.D. 1992, BAAS 24, 839
\ref Lloyd Evans T. 1993, private communication
\ref Maeder A. 1992, A\&A 264, 105
\ref Maeder A., Meynet G. 1993, A\&A 278, 406
\ref Nissen P.E., Edvardsson B. 1992, A\&A 261, 255
\ref Nomoto K., Thielemann F.-K., Yokoi K. 1984, ApJ 286, 644
\ref Nussbaumer H., Russca C. 1979, A\&A 72, 129
\ref Olofsson H., Eriksson K., Gustafsson B., Carlstr\"om U. 1993a, ApJS 87,
267
\ref Olofsson H., Eriksson K., Gustafsson B., Carlstr\"om U. 1993b, ApJS 87,
305
\ref Renzini A., Voli M. 1981, A\&A 94, 175
\ref Tomkin J., Lambert D.L. 1984, ApJ 279, 220
\ref Tomkin J., Lambert D.L., Edvardsson B., Gustafsson B., Nissen P.E. 1989,
A\&A 219, L15
\ref Tomkin J., Sneden C., Lambert D.L., 1986, ApJ 302, 415
\ref Wheeler J., Sneden C., Truran J.W. 1989, ARA\&A 27, 279
\endref
\end
\bye